\documentclass[epj,final]{svjour}
\smartqed  
\RequirePackage{graphicx}
\begin{document}

\title{Equivalence of the phenomenological Tsallis distribution to the transverse momentum distribution of $q$-dual statistics}

\author{A.S.~Parvan\inst{1}\inst{2}}

\institute{\inst{1}Bogoliubov Laboratory of Theoretical Physics, Joint Institute for Nuclear Research, Dubna, Russia \\ \inst{2}Department of Theoretical Physics, Horia Hulubei National Institute of Physics and Nuclear Engineering, Bucharest-Magurele, Romania}

\date{Received: date / Revised version: date}

\abstract{In the present work, we have found that the phenomenological Tsallis distribution (which nowadays is largely used to describe the transverse momentum distributions of hadrons measured in $pp$ collisions at high energies) is consistent with the basis of the statistical mechanics if it belongs to the $q$-dual nonextensive statistics instead of the Tsallis one. We have defined the $q$-dual statistics based on the $q$-dual entropy which was obtained from the Tsallis entropy under the multiplicative transformation of the entropic parameter $q\to 1/q$. We have found that the phenomenological Tsallis distribution is equivalent to the transverse momentum distribution of the $q$-dual statistics in the zeroth term approximation. Since the $q$-dual statistics is properly defined, it provides a correct link between the phenomenological Tsallis distribution and the second law of thermodynamics.
} 
\titlerunning{Equivalence of the phenomenological Tsallis distribution to}
\authorrunning{A.S.~Parvan}


\maketitle

Now, the phenomenological single-particle Tsallis distribution~\cite{Cleymans12,Cleymans12a} is used for the description of the experimental data on the transverse momentum distributions of hadrons created in the proton-proton and heavy-ion collisions at RHIC and LHC energies~\cite{Rybczynski14,Cleymans13,Azmi14,Marques13,Li14,Parvan17a,Biyajima06,Marques15,TsallisTaylor,TsallisRAA,BhattaCleMog,bcmmp}. The phenomenological Tsallis transverse momentum distribution for the Maxwell-Boltzmann statistics of particles introduced in Refs.~\cite{Cleymans12,Cleymans12a} has the form
\begin{eqnarray}\label{1}
\frac{d^{2}N}{dp_{T}dy} &=& \frac{gV}{(2\pi)^{2}} p_{T}  m_{T} \cosh y \nonumber \\ &\times& \left[1- (1-q) \frac{m_{T} \cosh y-\mu}{T} \right]^{\frac{q}{1-q}},
\end{eqnarray}
where $m_{T}$ is the transverse mass of particles. Initially, the phenomenological Tsallis transverse momentum distribution was proposed in another format (see Refs.~\cite{Bediaga00,Beck00})
\begin{eqnarray}\label{2}
\frac{1}{\sigma}\frac{d\sigma}{dp_{T}} &=& c p_{T} \int\limits_{0}^{\infty} dp_{L} \left[1- (1-q) \frac{\sqrt{p_{L}^{2}+m_{T}^{2}}}{T} \right]^{\frac{q}{1-q}},
\end{eqnarray}
where $p_{L}$ is the longitudinal momentum. However, further we will discuss only the function (\ref{1}) as it is widely used in high-energy physics.

In high-energy physics, the  phenomenological transverse momentum distribution (\ref{1}) is associated with the Tsallis statistics based on the Tsallis entropy~\cite{Tsal88}
\begin{equation}\label{3}
    S = \sum\limits_{i} \frac{p_{i}^{q}-p_{i}}{1-q}, \qquad  \sum\limits_{i} p_{i}=1,
\end{equation}
where $p_{i}$ is the probability of the $i$-th microscopic state of the system and $q\in\mathbf{R}$ is a real parameter taking values $0<q<\infty$.

In spite of the success of the function (\ref{1}) in the applications its connection with the basis of the statistical mechanics is still under the question. The phenomenological single-particle Tsallis distribution (\ref{1}) was obtained using the method of the maximization of generalized entropy of the ideal gas instead of the general Tsallis entropy (\ref{3}). This method is correct only for the usual Boltzmann-Gibbs statistics, but for the Tsallis statistics it is under the question~\cite{Parvan17}. Nevertheless, the correct connection of the distribution function (\ref{1}) with the Tsallis entropy (\ref{3}) was established in Ref.~\cite{Parvan17}. In that paper, it was rigorously demonstrated that the  phenomenological transverse momentum distribution (\ref{1}) corresponds to the zeroth term approximation of the Tsallis-2 statistics. In the case of massive particles, this result was confirmed in Ref.~\cite{Parvan19}. However, the Tsallis-2 statistics is improperly defined as the generalized expectation values of the thermodynamic quantities, $\langle A \rangle = \sum_{i} p_{i}^{q} A_{i}$, in this formalism are not consistent with probability normalization condition, $\sum_{i} p_{i}=1$ (see Ref.~\cite{Tsal98}). This means that the Tsallis-2 statistics disagrees with the probability theory and the second law of thermodynamics. Thus considering the phenomenological transverse momentum distribution (\ref{1}) as a single-particle distribution function of the Tsallis statistics based on the Tsallis entropy (\ref{3}) is a serious lack, because in this case  the distribution (\ref{1}) is obtained on the base of the erroneous general premises.

In the present study we demonstrate that this problem of connection of the phenomenological transverse momentum distribution (\ref{1}) with the basis of the statistical mechanics can be uniquely solved by introducing the new nonextensive statistics on the base of the $q$-dual entropy instead of the Tsallis entropy (\ref{3}).

{\it $q$-dual statistics}. The $q$-dual entropy is obtained from the Tsallis entropy (\ref{3}) by the multiplicative transformation $q\to 1/q$. Then, we have
\begin{equation}\label{4}
    S = q\sum\limits_{i} \frac{p_{i}^{1/q}-p_{i}}{q-1},
\end{equation}
where $q\in\mathbf{R}$ is a real parameter taking values $0<q<\infty$. In the Gibbs limit $q\to 1$, the entropy (\ref{4}) recovers the Boltzmann-Gibbs-Shannon entropy, $S=-\sum_{i} p_{i} \ln p_{i}$.

The statistical mechanics based on this entropy is defined by Eq.~(\ref{4}) with the probabilities $p_{i}$ of the microstates of the system normalized to unity
\begin{equation}\label{5}
    \varphi=\sum\limits_{i} p_{i} - 1 = 0
\end{equation}
and by the standard expectation values
\begin{equation}\label{6}
   \langle A \rangle = \sum\limits_{i} p_{i} A_{i}.
\end{equation}
Here, the definition of the expectation values (\ref{6}) is consistent with probability normalization constraint (\ref{5}). Let us consider the grand canonical ensemble. The thermodynamic potential $\Omega$ of the grand canonical ensemble is
\begin{eqnarray}\label{7}
 \Omega &=& \langle H \rangle -TS-\mu \langle N \rangle \nonumber \\
  &=&  \sum\limits_{i}  p_{i} \left[E_{i}-\mu N_{i} - T q \frac{p_{i}^{\frac{1}{q}-1}-1}{q-1}\right],
\end{eqnarray}
where $\langle H \rangle=\sum_{i}  p_{i} E_{i}$ is the mean energy of the system, $\langle N \rangle=\sum_{i}  p_{i} N_{i}$ is the mean number of particles of the system, and $E_{i}$ and $N_{i}$ are the energy and the number of particles, respectively, in the $i$-th microscopic state of the system.

The unknown probabilities $\{p_{i}\}$ are obtained from the second law of thermodynamics (the principle of maximum entropy). In the grand canonical ensemble they are found from the constrained local extrema of the thermodynamic potential (\ref{7}) by the method of the Lagrange multipliers (see, for example, Refs.~\cite{Jaynes2,Parvan2015,Krasnov}):
\begin{eqnarray}\label{8}
 \Phi &=& \Omega - \lambda \varphi,  \\ \label{9}
  \frac{\partial \Phi}{\partial p_{i}} &=& 0,
\end{eqnarray}
where $\Phi$ is the Lagrange function and $\lambda$ is an arbitrary real constant. Substituting Eqs.~(\ref{5}) and (\ref{7}) into Eqs.~(\ref{8}), (\ref{9}) and using Eq.~(\ref{5}) again, we obtain the normalized equilibrium probabilities of the grand canonical ensemble of the $q$-dual statistics as
\begin{equation}\label{10}
p_{i} = \left[1+(1-q)\frac{\Lambda-E_{i}+\mu N_{i}}{T}\right]^{\frac{q}{1-q}}
\end{equation}
and
\begin{equation}\label{11}
    \sum\limits_{i} \left[1+(1-q)\frac{\Lambda-E_{i}+\mu N_{i}}{T}\right]^{\frac{q}{1-q}}=1,
\end{equation}
where $\Lambda\equiv \lambda-T$ and $\partial E_{i}/\partial p_{i}=\partial N_{i}/\partial p_{i}=0$. In the Gibbs limit $q\to 1$, the probability $p_{i}=\exp[(\Lambda-E_{i}+\mu N_{i})/T]$, where $\Lambda=-T\ln Z$ is the thermodynamic potential of grand canonical ensemble and $Z=\sum_{i} \exp[-(E_{i} - \mu N_{i})/T]$ is the partition function.

Substituting Eq.~(\ref{10}) into Eq.~(\ref{6}), we obtain the statistical averages of the $q$-dual statistics in the grand canonical ensemble as
\begin{equation}\label{12}
   \langle A \rangle = \sum\limits_{i} A_{i} \left[1+(1-q)\frac{\Lambda-E_{i}+\mu N_{i}}{T}\right]^{\frac{q}{1-q}},
\end{equation}
where the norm function $\Lambda$ is the solution of Eq.~(\ref{11}). Note that the quantities (\ref{10})--(\ref{12}) of the $q$-dual statistics exactly recover the corresponding quantities of the Tsallis-1 statistics under the multiplicative transformation of the entropic parameter $q\to 1/q$ (see Ref.~\cite{Parvan2015}).

Let us rewrite the norm equation (\ref{11}) and the statistical averages (\ref{12}) in the integral representation in the case when the parameter $q>1$. Using the integral representation of the Gamma-function~\cite{Prato}, we can rewrite the norm equation (\ref{11}) for $q>1$ in the following form
\begin{equation}\label{13}
 \frac{1}{\Gamma\left(\frac{q}{q-1}\right)} \int\limits_{0}^{\infty} t^{\frac{1}{q-1}} e^{-t\left[1+(1-q)\frac{\Lambda-\Omega_{G}\left(\beta'\right)}{T}\right]} dt =1
\end{equation}
or
\begin{eqnarray}\label{14}
\sum\limits_{n=0}^{\infty} \frac{1}{n!\Gamma\left(\frac{q}{q-1}\right)} \int\limits_{0}^{\infty} t^{\frac{1}{q-1}}  e^{-t\left[1+(1-q)\frac{\Lambda}{T}\right]} &&  \nonumber \\  \left(-\beta'\Omega_{G}\left(\beta'\right)\right)^{n} dt &=& 1,
\end{eqnarray}
where $\beta'=t(q-1)/T$ and
\begin{eqnarray}\label{15}
  \Omega_{G}\left(\beta'\right) &=& -\frac{1}{\beta'} \ln Z_{G}\left(\beta'\right), \\ \label{16}
  Z_{G}\left(\beta'\right) &=& \sum\limits_{i} e^{-\beta'(E_{i}-\mu N_{i})}.
\end{eqnarray}
The statistical averages (\ref{12}) in the integral representation for $q>1$ can be rewritten as
\begin{eqnarray}\label{17}
  \langle A \rangle &=& \frac{1}{\Gamma\left(\frac{q}{q-1}\right)} \int\limits_{0}^{\infty}  t^{\frac{1}{q-1}} e^{-t\left[1+(1-q)\frac{\Lambda-\Omega_{G}\left(\beta'\right)}{T}\right]} \langle A \rangle_{G}\left(\beta'\right) dt   \nonumber \\ &=& \sum\limits_{n=0}^{\infty} \frac{1}{n!\Gamma\left(\frac{q}{q-1}\right)} \int\limits_{0}^{\infty} t^{\frac{1}{q-1}}  e^{-t\left[1+(1-q)\frac{\Lambda}{T}\right]}  \nonumber \\  &\times& \left(-\beta'\Omega_{G}\left(\beta'\right)\right)^{n} \langle A \rangle_{G}\left(\beta'\right) dt,
\end{eqnarray}
where
\begin{equation}\label{18}
  \langle A \rangle_{G}\left(\beta'\right) =\frac{1}{Z_{G}\left(\beta'\right)} \sum\limits_{i} A_{i} e^{-\beta'(E_{i}-\mu N_{i})}.
\end{equation}
Equation~(\ref{17}) links the statistical averages of the $q$-dual statistics with the corresponding statistical averages (\ref{18}) of the Boltzmann-Gibbs statistics.

{\it Transverse momentum distribution}. Let us consider the relativistic ideal gas with the Fermi-Dirac $(\eta=1)$, Bose-Einstein $(\eta=-1)$ and Maxwell-Boltzmann $(\eta=0)$ statistics of particles in the grand canonical ensemble. The transverse momentum distribution of particles is related to the mean occupation numbers of the ideal gas as
\begin{equation}\label{19}
  \frac{d^{2}N}{dp_{T}dy} = \frac{V}{(2\pi)^{3}} \int\limits_{0}^{2\pi} d\varphi p_{T} \varepsilon_{\mathbf{p}} \ \sum\limits_{\sigma} \langle n_{\mathbf{p}\sigma}\rangle,
\end{equation}
where $\varepsilon_{\mathbf{p}}=m_{T} \cosh y$ is the single-particle energy, $p_{T}$ and $y$ are the transverse momentum and rapidity variables, respectively, and $m_{T}=\sqrt{p_{T}^{2}+m^{2}}$. The mean occupation numbers $\langle n_{\mathbf{p}\sigma}\rangle$ are calculated from Eq.~(\ref{17}) using the mean occupation numbers of the ideal gas of the Boltzmann-Gibbs statistics:
\begin{equation}\label{20}
  \langle n_{\mathbf{p}\sigma} \rangle_{G}\left(\beta'\right) = \frac{1}{e^{\beta' (\varepsilon_{\mathbf{p}}-\mu)}+\eta}.
\end{equation}
Considering Eqs.~(\ref{17}), (\ref{19}) and (\ref{20}), we obtain
\begin{eqnarray}\label{21}
  \frac{d^{2}N}{dp_{T}dy} &=& \frac{gV}{(2\pi)^{2}} p_{T}  m_{T} \cosh y  \sum\limits_{n=0}^{\infty}  \frac{1}{n!\Gamma\left(\frac{q}{q-1}\right)} \nonumber \\ &\times&
  \int\limits_{0}^{\infty} t^{\frac{1}{q-1}} e^{-t\left[1+(1-q)\frac{\Lambda}{T}\right]} \frac{\left(-\beta'\Omega_{G}\left(\beta'\right)\right)^{n}}{e^{\beta' (m_{T} \cosh y-\mu)}+\eta} dt, \;\;\;\;
\end{eqnarray}
where
\begin{equation}\label{22}
  -\beta'\Omega_{G}\left(\beta'\right)= \sum\limits_{\mathbf{p},\sigma} \ln \left[1+\eta e^{-\beta' (\varepsilon_{\mathbf{p}}-\mu)} \right]^{\frac{1}{\eta}}
\end{equation}
is the thermodynamic potential for the relativistic ideal gas of the Boltzmann-Gibbs statistics of microstates and $\varepsilon_{\mathbf{p}}=\sqrt{\mathbf{p}^2+m^{2}}$. The norm function $\Lambda$ for $\eta=-1,0,1$ is calculated from the norm equation~(\ref{14}) using Eq.~(\ref{22}).

{\it Maxwell-Boltzmann statistics of particles}. In the limit $\eta\to 0$, the thermodynamic potential of the ideal gas of the Boltzmann-Gibbs statistics (\ref{22}) can be written as
\begin{equation}\label{23}
  \Omega_{G}\left(\beta'\right)= - \frac{gV}{2\pi^{2}} \frac{m^{2}}{\beta'^{2}} e^{\beta'\mu} K_{2}\left(\beta'm\right),
\end{equation}
where $K_{\nu}(z)$ is the modified Bessel function of the second kind. Substituting Eq.~(\ref{23}) into Eq.~(\ref{14}), we obtain the norm equation for the Maxwell-Boltzmann statistics of particles as
\begin{eqnarray}\label{24}
 && \sum\limits_{n=0}^{\infty} \frac{\omega^{n}}{n!} \frac{1}{\Gamma\left(\frac{q}{q-1}\right)} \int\limits_{0}^{\infty} t^{\frac{1}{q-1}-n} e^{-t\left[1+(1-q)\frac{\Lambda+\mu n}{T}\right]} \nonumber \\ && \qquad \left(K_{2}\left(\frac{t(q-1)m}{T} \right)\right)^{n} dt = 1,
\end{eqnarray}
where
\begin{equation}\label{25}
  \omega=\frac{gVTm^{2}}{2\pi^{2}} \frac{1}{q-1}.
\end{equation}

Substituting Eq.~(\ref{23}) into Eq.~(\ref{21}) and taking the limit $\eta=0$, we obtain the transverse momentum distribution for the Maxwell-Boltzmann statistics of particles as
\begin{eqnarray}\label{26}
  \frac{d^{2}N}{dp_{T}dy} &=& \frac{gV}{(2\pi)^{2}} p_{T}  m_{T} \cosh y  \sum\limits_{n=0}^{\infty} \frac{\omega^{n}}{n!} \frac{1}{\Gamma\left(\frac{q}{q-1}\right)} \nonumber \\
   &\times&   \int\limits_{0}^{\infty} t^{\frac{1}{q-1}-n}
 e^{-t\left[1+(1-q)\frac{\Lambda-m_{T} \cosh y+\mu(n+1)}{T}\right]} \nonumber \\ &\times& \left(K_{2}\left(\frac{t(q-1)m}{T} \right)\right)^{n} dt.
\end{eqnarray}

{\it Zeroth term approximation}. Let us find the transverse momentum distribution of particles in the zeroth term approximation, which was introduced in Ref.~\cite{Parvan17}. In this approximation we retain only the zeroth term $(n=0)$ in the series expansion. Taking $n=0$ in Eq.~(\ref{14}), we obtain that the norm function $\Lambda=0$. Substituting $\Lambda=0$ into Eq.~(\ref{21}) and considering only the zeroth term and the equation
\begin{equation}\label{27}
\frac{1}{e^{x}+\eta} = \sum\limits_{k=0}^{\infty} (-\eta)^{k} e^{-x(k+1)},
\end{equation}
where $|e^{-x}|<1$, we obtain the transverse momentum distribution in the zeroth term approximation as
\begin{eqnarray}\label{28}
\frac{d^{2}N}{dp_{T}dy} &=& \frac{gV}{(2\pi)^{2}} p_{T}  m_{T} \cosh y \nonumber \\ &\times& \sum\limits_{k=0}^{\infty} (-\eta)^{k}  \left[1- (k+1) (1-q) \frac{m_{T} \cosh y-\mu}{T} \right]^{\frac{q}{1-q}}  \nonumber \\ &&   \mathrm{for} \quad \eta=-1,0,1   \;\;\;\;\;\;\;\;\;\;
\end{eqnarray}
or
\begin{eqnarray}\label{29}
\frac{d^{2}N}{dp_{T}dy} &=& \frac{gV}{(2\pi)^{2}} p_{T}  m_{T} \cosh y  \nonumber \\ &\times& \left[1 - (1-q) \frac{m_{T} \cosh y-\mu}{T} \right]^{\frac{q}{1-q}} \nonumber \\ &&   \mathrm{for} \quad \eta=0.
\end{eqnarray}
The transverse momentum distribution (\ref{29}) exactly coincides with the phenomenological Tsallis distribution (\ref{1}) introduced in~\cite{Cleymans12,Cleymans12a} (see Eq.~(56) in Ref.~\cite{Cleymans12}). Thus, the phenomenological Tsallis distribution (\ref{1}) for the Maxwell-Boltzmann statistics of particles exactly corresponds to the zeroth term approximation of the $q$-dual nonextensive statistics based on the entropy (\ref{4}). Moreover, the transverse momentum distribution (\ref{29}) of the $q$-dual nonextensive statistics recovers the transverse momentum distribution of the Tsallis-1 statistics in the zeroth term approximation under the transformation of the entropic parameter $q\to 1/q$~\cite{Parvan17,Parvan19}. However, the phenomenological Tsallis distributions for the quantum (Fermi-Dirac and Bose-Einstein) statistics of particles introduced in Ref.~\cite{Cleymans12} do not recover the transverse momentum distributions in the zeroth term approximation of both the $q$-dual nonextensive statistics (cf. Eqs.~(31) and (33) of Ref.~\cite{Cleymans12} along with Eq.~(\ref{28}) of the present paper) and the Tsallis statistics~\cite{Parvan19}.

To summarize, we have introduced the $q$-dual entropy which is defined from the Tsallis entropy by the multiplicative transformation of the entropic parameter $q\to 1/q$. The $q$-dual nonextensive statistics based on this statistical entropy and consistently defined expectation values of thermodynamic quantities was formulated in the framework of the grand canonical ensemble. The thermodynamic potential was obtained from the fundamental thermodynamic potential by the Legendre transform. The probabilities of microstates of the system were derived from the second law of thermodynamics (the Jaynes maximum entropy principle) by the constrained local extrema of the statistical thermodynamic potential. The norm equation  and statistical averages were expressed analytically in terms of the corresponding Boltzmann-Gibbs quantities using the integral representation and the exponential function containing the thermodynamic potential was expanded into the series. We have obtained the exact analytical formulae for the transverse momentum distributions of the relativistic massive particles following the Bose-Einstein, Fermi-Dirac and Maxwell-Boltzmann statistics in the framework of $q$-dual statistics in the grand canonical ensemble.

In the present study, we have found that the phenomenological Tsallis distribution for the Maxwell - Boltzmann statistics of particles introduced in Refs.~\cite{Cleymans12,Cleymans12a} is equivalent to the transverse momentum distribution of the $q$-dual nonextensive statistics in the zeroth term approximation. This demonstrates that the correct link between the phenomenological Tsallis distribution and the fundamental theory of the statistical mechanics is provided by the $q$-dual entropy instead of the Tsallis one. The $q$-dual statistics is properly defined in comparison with the Tsallis-2 statistics as the expectation values of this formalism are consistent with probability normalization condition. Thus we have found that the well-known phenomenological single-particle Tsallis distribution is thermodynamically founded only if it belongs to the $q$-dual nonextensive statistics instead of the Tsallis one.

\begin{acknowledgement}
This work was supported in part by the joint research projects of JINR and IFIN-HH.
\end{acknowledgement}

\end{document}